%% file: MaxDemon-arXiv-v2.tex
\documentclass[%
 reprint,
superscriptaddress,
showpacs,preprintnumbers,
 amsmath,amssymb,
 aps,
pra,
]{revtex4-1}

\usepackage{graphicx}% Include figure files
\usepackage{dcolumn}% Align table columns on decimal point
\usepackage{bm}% bold math
\usepackage{color}
\usepackage{psfrag}
\newcommand{\refeq}[1]{Eq.~\eqref{#1}}

\newtheorem{prop}{Proposition}

\begin{document}

\title{Maximum work extraction and implementation costs \\ for non-equilibrium Maxwell's demons}

\author{Henrik Sandberg}
\affiliation{Department of Automatic Control, KTH Royal Institute of Technology, Stockholm, Sweden}

\author{Jean-Charles Delvenne}
\affiliation{ICTEAM and CORE, Universit{\'e} catholique de Louvain, Louvain-la-Neuve, Belgium}

\author{Nigel J. Newton}
\affiliation{School of Computer Science and Electronic Engineering, University of Essex, Colchester, UK}

\author{Sanjoy K. Mitter}
\affiliation{Laboratory for Information and Decision Systems, MIT, Cambridge, Massachusetts, USA}

\date{July 22, 2014}

\begin{abstract}
In this theoretical study, we determine the maximum amount of work extractable in finite time by a demon performing continuous
measurements on a quadratic Hamiltonian system subjected to thermal fluctuations, in terms of the information extracted from the system. This is in contrast to many recent studies that focus on demons' maximizing the extracted work over received information, and operate close to equilibrium.
 The maximum work demon is found to apply a high-gain continuous feedback using a Kalman-Bucy estimate of the system state. A simple and concrete electrical implementation of the feedback protocol is proposed, which allows for analytic expressions of the flows of energy and entropy inside the demon. This let us show that any implementation of the demon must necessarily include an external power source, which we prove both from classical thermodynamics arguments and from a version of Landauer's memory erasure argument extended to non-equilibrium linear systems.
\end{abstract}


\maketitle

Ever since Maxwell \cite{maxwell} put forward the idea of an abstract being (a demon) apparently able to break the second law of thermodynamics, it has served as a great source of inspiration and helped to establish important connections between statistical physics and information theory. See, for example, \cite{szilard,landauer61,bennett,penrose,maxwell2}.
In the original version, the demon operates a trapdoor between two heat baths, such that a seemingly counterintuitive heat flow is established. Today, more generally, devices that are able to extract work from a single heat bath by rectifying thermal fluctuations are also called `Maxwell's demons' \cite{jarzynski+13}. Several schemes detailing how the demon could apparently break the second law have been proposed, for example Szilard's heat engine \cite{szilard}. More recent schemes are presented in \cite{horowitz+10,ueda+12,jarzynski+13,esposito+13}, and \cite{jarzynski+12,barao+13} where measurement errors are also accounted for.

A classical expression of the second law states that the maximum (average) work extractable from a system
in contact with a single thermal bath cannot exceed
the free energy decrease between the system's initial and final equilibrium states.
However, as illustrated by Szilard's heat engine, it is possible to break this bound under the assumption of additional information available to the work-extracting agent. To account for this possibility, the second law can be generalized to include transformations using \emph{feedback control} \cite{sagawa+10,fujitani+10,horowitz+10,abreu+11,sagawa+12,sagawa+13}. In particular, in \cite{sagawa+12} it is shown that under feedback control, the extracted work $W$ must satisfy
\begin{equation}
W \leq kT I_c,
\label{eq:2ndlaw}
\end{equation}
where $k$ is Boltzmann's constant, $T$ is the temperature of the bath, and $I_c$ is the so called \emph{transfer entropy} from the system to the measurement. Note that in~\refeq{eq:2ndlaw} we have assumed there is no free energy decrease from the initial to final state. Related generalizations of the second law are stated in \cite{hasegawa+10,esposito+11,deffner+13}. It is possible to construct feedback protocols that saturate \refeq{eq:2ndlaw} using reversible and quasi-static transformations \cite{horowitz+11,sagawa+12,horowitz+13}. Reversible feedback protocols may be optimal in terms of making \refeq{eq:2ndlaw} tight, but they are also infinitely slow, and in \cite{schmiedl+07,abreu+11,bauer+12} some related finite-time problems are addressed.

The first contribution of this paper is to state an explicit \emph{finite-time} counterpart to \refeq{eq:2ndlaw},
characterizing the maximum work extractable using feedback control, in terms of the transfer entropy.
To explain our result, consider a system modeled by an overdamped Langevin equation. We show that the maximum amount of extractable work over a duration $t$, $W_{\max}(t)$, can be expressed by the integral
 \begin{equation}
W_{\max}(t)  = k \int_0^t T_{\min}  \dot{I}_c \,dt' \leq kT I_c(t).
\label{eq:mainres}
\end{equation}
Here $T_{\min}(t)$ has an interpretation as the lowest achievable system temperature after $t$ time units of
continuous feedback control, assuming an equilibrium initially $T_{\min}(0)=T$.
Since $T_{\min}(t)\leq T$, for all $t$, the upper bound in \refeq{eq:mainres} follows trivially, implying \refeq{eq:2ndlaw}.
The transfer entropy $I_c(t)$ measures the useful amount of information transmitted to the controller
from the partial observations in the time interval $[0,t]$. Therefore, every bit of transfer entropy, if optimally
exploited, allows us to retrieve between $kT_{\min}\ln 2$ and $kT \ln 2$ units of work.
We furthermore provide a novel expression for the transfer entropy $I_c(t)$, applicable to a large class of systems in both continuous and in discrete time. In particular, the new expression yields closed-form solutions of the transfer entropy and  shows its independence of the applied feedback law.

For systems of dimension higher than one satisfying linear dynamics, e.g., systems with quadratic Hamiltonians in contact with a heat bath, we show that
\begin{equation}
W_{\max}(t) \simeq k \int_0^t T_{\min}  \dot{I}_c \,dt',
\label{eq:mainres2}
\end{equation}
with asymptotic equality as $t\rightarrow \infty$, i.e., in non-equilibrium steady state (NESS).
Furthermore, it always holds that $W_{\max}(t)\leq kT I_c(t)$ and the second law is validated for all finite time intervals.
A quadratic Hamiltonian is a common and reasonable assumption for a system excited by thermal fluctuations of moderate temperature around a minimum-energy state.

Our second contribution is to use control theory to characterize and interpret the feedback protocol the demon should apply to reach the upper limit $W_{\max}(t)$. The protocol is a linear feedback law based on the optimal estimate of the system state, which can be recursively computed using the so-called Kalman-Bucy filter.
The found feedback law also offers a simple electrical implementation. A proper physical implementation is also shown to require an external work supply to maintain the noise on the wires at an acceptable level. The cost of this noise suppressing mechanism can be evaluated by standard thermodynamic arguments, or through a non-equilibrium extension of Landauer's memory erasure principle.

The major difference of this paper compared to \cite{allahverdyan+09,sagawa+10,fujitani+10,horowitz+10,abreu+11,hasegawa+10,esposito+11,
sagawa+12,deffner+13,sagawa+13}, for example, is the focus on demons extracting work at maximum rate, rather than focusing on bounding or maximizing the extracted work per bit of received information (`information efficiency'). A large information efficiency is associated with slow and reversible work extraction close to equilibrium, and so our study naturally leads to demons operating at non-equilibrium.
Most practically relevant feedback systems, in engineering and in biology, by design operate at non-equilibrium, and we argue that characterizing limits on their energy conversion rates is a problem of major importance.
Perhaps surprisingly, we find that maximum work demons obey several insightful relations, including \refeq{eq:mainres}, information efficiency, and implementation costs, that are independent on many of the precise details of the demons.
Another difference of our work is that we exploit some stochastic calculus techniques, well known in control engineering, but less frequently used in the physics literature. This most of the time lets us work directly with continuous-time models, yielding simple closed-form solutions of $I_c(t)$ and $T_{\min}(t)$, for example, instead of using discrete-time models and limiting arguments \cite{sagawa+12,sagawa+13,barato+13,diana+13}. An advantage of our continuous-time approach is that it immediately suggests how to physically implement the found feedback protocols. Traditional controller design in engineering science is often done in a continuous-time setting \cite{bechhoefer05,astrom+08}.
The papers \cite{aurell+11,aurell+12,delvenne+13} do employ continuous-time feedback as well, but exact system state knowledge by the controller is assumed there, which significantly simplifies the studied optimal-control problems. The paper \cite{munakata+13} also studies continuous-time feedback with measurement errors, but does not characterize information flows and maximum work bounds of the type in \refeq{eq:mainres}.

The structure of the paper is as follows. In Section~\ref{sec:sysmodel}, we introduce an electrical system that is modeled by an overdamped Langevin equation. This is the example system used throughout the paper, although we show in Appendix~\ref{sec:multidim} how to generalize the results to higher-dimensional systems. In Section~\ref{sec:demonmodel}, we introduce the model of the demons, and the Kalman-Bucy filter used for their implementation. In Section~\ref{sec:infoflow}, we give a novel characterization of the transfer entropy $I_c(t)$, and use it to
obtain the maximum work relation in \refeq{eq:mainres}. The derivation of the transfer entropy is found in Appendix~\ref{sec:TE}. In Section~\ref{sec:maxefficient}, we present some alternative demon schemes, to challenge and give further insight to \refeq{eq:mainres}.
In Section~\ref{sec:impl}, we show how the demons can be built using simple electrical components. We also compute their implementation costs. For this, an expression for the information rate to the memory is needed, which is derived in Appendix~\ref{sec:inforate}.
We conclude the paper in Section~\ref{sec:disc} with a summary and discussion of the results.

\begin{figure}[tb]
\centering
\includegraphics[width=1.00\hsize]{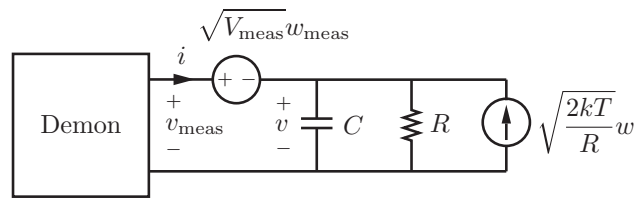}\\
  \caption{The demon (the feedback controller) connected to a capacitor, a heat bath of temperature $T$, and a measurement
  noise source of intensity $V_\text{meas}$. The demon may choose the current $i$ freely, and has access to the noisy voltage measurement $v_\text{meas}$.}\label{fig:RCcircuit}
\end{figure}

\section{System model}
\label{sec:sysmodel}
The system we first consider is an electric capacitor $C$, a resistor $R$ with thermal noise (the heat bath), and a feedback controller (the demon) with access to noisy voltage measurements, see Fig.~\ref{fig:RCcircuit}.
The resistor is subjected to Johnson-Nyquist noise \cite{johnson28,nyquist28}.
The circuit is modeled by an overdamped Langevin equation
\begin{equation}
\begin{aligned}
\tau \dot v & = - v + R i + \sqrt{2kTR}w, & \, \langle v(0)\rangle & = 0, \\
v_\text{meas} & = v + \sqrt{V_\text{meas}} w_\text{meas}, & \, \langle v(0)^2 \rangle &= \frac{kT}{C},
\end{aligned}
\label{eq:RCcircuit}
\end{equation}
 with $v(0)$ Gaussian, $w$ and $w_\text{meas}$ uncorrelated Gaussian white noise ($\langle w(t)w(t')\rangle=\langle w_\text{meas}(t)w_\text{meas}(t')\rangle = \delta(t-t')$),
 $V_\text{meas}$ the intensity of the measurement noise, and $\tau=RC$ being the time constant of the open circuit. The measurement noise $\sqrt{V_\text{meas}} w_\text{meas}$ can be thought of as the Johnson-Nyquist noise of the wire between the capacitor and the demon, whose resistance for simplicity is incorporated in the demon.
The heat flow to the capacitor is $\dot Q$ and the work-extraction rate of the demon is $\dot W$, and satisfy the first law of thermodynamics,
\begin{equation}
\dot U  = \dot Q - \dot W
\label{eq:1stlawA}
\end{equation}
where
\begin{equation}
\begin{aligned}
 U & = \frac{1}{2} C \langle v^2 \rangle  \equiv \frac{1}{2}kT_C, \\
 \dot Q  & = \frac{k}{\tau}(T-T_C), \quad \dot W   = - \langle v i\rangle.
\end{aligned}
\label{eq:heatflows}
\end{equation}
We denote the \emph{effective}
instantaneous temperature (`kinetic temperature') of the capacitor by $T_C$, and its internal energy by $U$. For detailed derivations of Eqs.~\eqref{eq:1stlawA}--\eqref{eq:heatflows}, see \cite{delvenne+13}. Furthermore, we assume the
capacitor initially is in thermal equilibrium with the heat bath, i.e., $T_C(0)=T$. Just as in \cite{abreu+11}, we can justify
calling $T_C(t)$ a temperature since it appears in a Fourier-like heat conduction law (see $\dot Q$).
 Also, since our applied controls will maintain a Gaussian distribution of $v$, $T_C(t)$ will be the true temperature of the capacitor if
  it were to be disconnected from all the other elements at time $t$.
The voltage $v_\text{meas}$ is the measurement that supplies the demon with information, and can be seen as a noisy measurement of the fluctuating capacitor voltage $v$. We will show how a demon can optimally control the work extraction
 by carefully exploiting the measurements $v_\text{meas}$ and properly choosing the injected current $i$. Intuitively, the demon can create a positive work rate $\dot W$ if it chooses $i<0$ when it correctly estimates $v>0$, and vice versa. But how the demon should estimate
$v$, and how to optimally choose $i$ may be less obvious.

If we know the trajectory of the effective temperature $T_C$, it is from Eqs.~\eqref{eq:1stlawA}--\eqref{eq:heatflows} possible
to solve for the amount of extracted work,
\begin{equation}
W(t)= \int_0^t \frac{k}{\tau}(T-T_C) \, dt' + \frac{1}{2}k(T-T_C(t)).
\label{eq:W}
\end{equation}
In particular, if we can characterize a lower bound on the effective temperature under all allowed controls,
$T_{\min}(t')\leq T_C(t')$ for $0\leq t' \leq t$, we get an upper bound on the
work that a demon can extract,
\begin{equation}
W_{\max}(t):= \int_0^t \frac{k}{\tau}(T-T_{\min}) \, dt' + \frac{1}{2}k(T-T_{\min}(t)),
\label{eq:Wmax}
\end{equation}
so that $W(t) \leq W_{\max}(t)$. In the following, we characterize $T_{\min}$, and thereby $W_{\max}$, using optimal control theory.

\section{Demon model and optimal continuous-time feedback}
\label{sec:demonmodel}
Optimal control theory \cite{Astrom} teaches how to compute $T_{\min}$, and to characterize the corresponding feedback law. In particular, for linear systems the \emph{separation principle} \cite{wonham68} says we can achieve the goal in two steps:
First, we should continuously and optimally estimate the voltage $v(t)$ of the capacitance, given the available
measurements $(v_\text{meas})_0^t \equiv \{v_\text{meas}(t'), \quad 0\leq t' \leq t \}$. Second, we should continuously use the found optimal estimate to update the current $i(t)$ using a suitable linear feedback law.

The best possible estimate $\hat v(t)$ of $v(t)$, given the measurement trajectory $(v_\text{meas})_0^t$,
can be recursively constructed by the celebrated \emph{Kalman-Bucy filter} \cite{bucy+68}, which leads to a minimum variance estimation error \cite{Astrom} and exploits as much of the information contained in $v_\text{meas}$ as is possible \cite{Mitter+05}. The Kalman-Bucy filter for \refeq{eq:RCcircuit} is given by
\begin{equation}
\tau \frac{d}{dt} \hat v  = - \hat v + R i
+ \frac{\sigma T_{\min}}{2T}\left( v_\text{meas}-\hat v  \right), \quad \hat v(0)=0,
\label{eq:KalmanBucy}
\end{equation}
where $T_{\min}$ solves the filter Riccati equation over the time interval $[0,t]$,
\begin{equation}
\tau \dot T_{\min} = 2 (T -T_{\min}) - \frac{\sigma T_{\min}^2}{2 T}, \quad T_{\min}(0) = T,
\label{eq:riccati}
\end{equation}
and
\begin{equation}
\sigma \equiv \frac{2kTR}{V_\text{meas}}
\end{equation}
is  a fundamental adimensional characterization of the bath noise compared to the measurement noise.
A large value of $\sigma$ signifies a demon with access to high-quality measurements, and vice versa.
The initial conditions $\hat v(0)=0$ and $T_{\min}(0) = T$ reflect the fact that
the best unbiased estimate initially is zero, and that the demon knows the temperature of the bath.
The current $i$ in \refeq{eq:KalmanBucy} is identical to the current applied by the demon to the system, and
can be any well-behaved causal feedback control policy \cite{wonham68}. That is, $i(t)=f_t(v_\text{meas})$ for some functional $f_t$, which only depends on the measurements received until time $t$, i.e., $(v_\text{meas})_0^t$. Note that the Kalman-Bucy filter can be implemented online in a feedback
controller, since it causally depends on the measurement realization $v_\text{meas}$, and $T_{\min}$ can be solved for offline.

\begin{figure}[tb]
\centering
  \psfrag{t}[][]{$t$}
  \psfrag{Tmin}[][]{$T_{\min}$}
  \psfrag{T113}[][]{$\sigma=1,\,\tau=3$}
  \psfrag{T11}[][]{$\sigma=1,\,\tau=1$}
  \psfrag{T21}[][]{$\sigma=2,\,\tau=1$}
  \psfrag{dotWmax}[][]{$\dot W_{\max}$}
  \includegraphics[width=1.00\hsize]{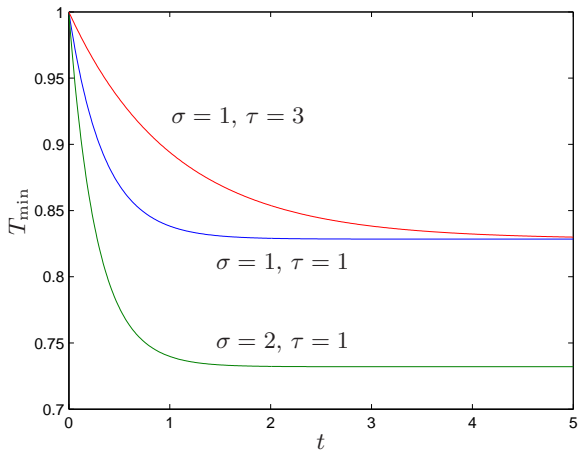}
  \caption{Transient effective temperature $T_\text{min}$ of a capacitor being optimally cooled by a demon. The capacitor starts in equilibrium ($T=1$), and then rapidly reaches a cooler NESS, whose level is solely determined by the measurement quality $\sigma$ and bath temperature $T$. In particular, a very accurate measurement ($\sigma$ large) leads to an effective temperature close to zero.}\label{fig:Tmin_vs_time}
\end{figure}

As we shall see, $T_{\min}$ solving \refeq{eq:riccati} is identical to the minimum possible effective temperature
$T_{\min}$ which determines $W_{\max}(t)$ in \refeq{eq:Wmax}.
The Riccati equation \refeq{eq:riccati} has a closed-form solution,
\begin{multline}
T_{\min}(t) = T_{\min}^\text{NESS} \\ + \dfrac{(T-T_{\min}^\text{NESS})e^{-2\gamma t/\tau}}{1 + \sigma(T-T_{\min}^\text{NESS})
(1-e^{-2\gamma  t/\tau})/(4 \gamma  T)}
\end{multline}
where $\gamma=\sqrt{1 + \sigma}>1$. From the solution it is seen NESS is approached monotonically and exponentially fast. Starting from $T$, $T_{\min}$ decreases exponentially, and monotonically, to a steady-state value
\begin{equation}
 T_{\min}^{\text{NESS}}  =  \frac{2T}{\sqrt{1 + \sigma} + 1}< T.
\label{eq:TminNESS}
\end{equation}
In the noisy measurement limit $\sigma \ll 1$, the NESS may reach the effective temperature $(1-\sigma/4)T$, slightly colder than $T$, while accurate measurements $\sigma \gg 1$ allows us to reach a low effective temperature $2T/\sqrt{\sigma}$.
The case $\sigma \approx 0$, and thus $T_{\min}^\text{NESS}\approx T$, will be of some interest in the following. Therefore we call this the \emph{poor measurement limit}. In particular, in this limit the capacitor will be close to equilibrium with the bath, even if the demon extracts work at the maximum possible rate.
In Fig.~\ref{fig:Tmin_vs_time}, three transient trajectories of $T_{\min}$ are shown. It is seen that NESS is reached in the time of order $\tau$.

Guided by optimal control theory and the separation principle,
we let the demon use the simple linear causal feedback
\begin{equation}
i(t') = - G \hat v(t'), \quad 0\leq t' \leq t,
\label{eq:feedback}
\end{equation}
where $0 \leq G < \infty$ is a fixed scalar feedback gain.
We may think of the feedback gain $G$ as the `conductance' of the demon: If the demon believes the voltage of the capacitor to be $\hat v$, it will admit the current $G\hat v$. If $v \approx \hat v$, the demon will indeed look like an electric load
of conductance close to $G$. While $G=0$ (open circuit) creates a demon that only (optimally) observes, $G \to \infty$ also removes energy from the capacitance at the highest possible rate, achieving the minimum effective temperature $T_{\min}$. This can be seen as follows:
 Inserting \refeq{eq:feedback} in \refeq{eq:KalmanBucy} we can compute the evolution of the variance $\hat V\equiv \langle \hat v^2 \rangle$ of the filter estimate as
\begin{equation}
\tau \frac{d}{dt} \hat V =  -2\left(1 + GR \right) \hat V + \frac{\sigma kT_{\min}^2}{2C T}, \quad \hat V (0)=0.
\label{eq:control_heat}
\end{equation}
We note that since $T_{\min}$ is bounded, $\hat V$ can be made arbitrarily close to zero by increasing the feedback gain $G$.
The Kalman-Bucy filter is variance-optimal, i.e., the variance of the estimation error
$\langle \Delta v^2 \rangle \equiv \langle [v - \hat v ]^2\rangle = k T_{\min}/C$ is the smallest possible \cite{Astrom}.
Then the estimation error $\Delta v$ is orthogonal to the estimate \cite{Astrom}, $\langle \hat v \Delta v\rangle = 0$, and therefore
\begin{equation}
 \frac{k T_{C}}{C} = \langle v^2 \rangle = \langle \hat v^2\rangle  + \langle \Delta v^2 \rangle = \hat V + \frac{k T_{\min}}{C}.
\end{equation}
 Since $T_{\min}$ is independent of $G$, and $\hat V$ can be made arbitrarily close to zero, we realize that the demon through its policy is cooling the capacitor and for all $t$,
\begin{equation}
T_C(t) \searrow T_{\min}(t) \quad \text{as} \quad  G\rightarrow \infty.
\end{equation}
This shows a demon should implement a Kalman-Bucy filter with a large (infinite) feedback gain $G$ to extract the work $W_{\max}$.

For a general feedback gain $G\geq 0$ in \refeq{eq:feedback}, the effective temperature of the capacitor will drop exponentially from $T_C(0)=T$ to
\begin{equation}
T_C^\text{NESS} = \frac{1}{1+GR} T + \frac{GR}{1+GR}T_{\min}^\text{NESS}.
\end{equation}
The corresponding NESS work-extraction rate can be shown to become
\begin{equation}
\dot W^\text{NESS} = \frac{k}{\tau}(T-T_{\min}^\text{NESS})\frac{GR}{1+GR}.
\label{eq:NESSwork}
\end{equation}
Thus the continuous feedback protocol in \refeq{eq:feedback} can realize any NESS work rate between $0$ and the maximum $\dot W_{\max}^\text{NESS} = \frac{k}{\tau}(T-T_{\min}^\text{NESS})$ by proper choice of gain $G$.

The above optimal controller can be generalized to any system with linear dynamics. Details are given in Appendix~\ref{sec:multidim}
for systems with quadratic Hamiltonians.

\section{Information flow and maximum work theorem}
\label{sec:infoflow}
To establish the maximum work theorem in \refeq{eq:mainres}, we need to quantify the information flow from
the uncertain part of the voltage $v$ to the measurement $v_\text{meas}$, under continuous feedback.
This is the \emph{transfer entropy}, as is explained in~\cite{sagawa+12}, for example. We show in Appendix~\ref{sec:TE} that
the appropriate continuous-time limit of the transfer entropy is \footnote{The rigorous definition in terms of integrated continuous-time stochastic processes is found in Appendix~\ref{sec:TE}.}
\begin{equation}
I_c(t) = I((v(0),(w)_0^t);(v_\text{meas})_0^t).
\label{eq:Icdef}
\end{equation}
This is the \emph{mutual information} between the uncertain initial voltage $v(0)$ and noise trajectory $w$ from the bath, and the measurement trajectory $v_\text{meas}$. Mutual information~\cite{coverthomas} between two stochastic variables $\xi$ and $\theta$
is as usual defined as
\begin{equation}
\begin{aligned}
I(\theta;\xi) \equiv \int  \ln \left(\frac{d \mathbb{P}_{\theta \xi}}{d(\mathbb{P}_{\theta} \otimes \mathbb{P}_{\xi})}\right) d\mathbb{P}_{\theta \xi }\geq 0,
\end{aligned}
\label{eq:mutualinfodef}
\end{equation}
and is equal to the amount the (differential) Shannon entropy of $\xi$ decreases with knowledge of $\theta$, and vice versa.
Here $\mathbb{P}_{\theta \xi},\, \mathbb{P}_{\theta}$, and $\mathbb{P}_{\xi}$ are joint and marginal probability measures of the stochastic variables $\theta$ and $\xi$.
We prove in Appendix~\ref{sec:TE} that the transfer entropy in fact has the following explicit form:
\begin{equation}
I_c(t) = \frac{\sigma}{4\tau} \int_0^t \frac{T_{\min}}{T} \, dt'.
\label{eq:mutualinfo}
\end{equation}
Note that $I_c$ \emph{does not} otherwise depend on the details of the demon, for example the feedback gain $G$, as is further discussed in Appendix~\ref{sec:TE}.

It now follows from Eqs.~\eqref{eq:Wmax}, \eqref{eq:riccati}, and \eqref{eq:mutualinfo}
that the maximum extracted work must satisfy
\begin{equation}
\begin{aligned}
W_{\max}(t) & = \int_0^t \frac{k}{\tau}(T-T_{\min}) \, dt' + \frac{1}{2}k(T-T_{\min}(t)) \\
& = \int_0^t \frac{\sigma kT_{\min}^2}{4 \tau  T} \, dt' =  k \int_0^t T_{\min} \dot{I}_c\, dt',
\end{aligned}
\end{equation}
which proves the equality in \refeq{eq:mainres}. The inequality trivially follows since $T_{\min}\leq T$.
As is shown in Appendix~\ref{sec:multidim}, for multi-dimensional systems we only have an asymptotic equality: As soon as the system satisfies an equipartition condition (true at all times in the one-dimensional case, and in any dimension when a NESS is reached) the equality holds.

The expressions for $I_c$ and $W_{\max}$ provide interesting insights concerning information and
work flow in the feedback loop. Since $T_{\min}$ decreases monotonically,
the transfer entropy rate $\dot{I}_c$ is largest just when the measurement and feedback control start, and then decreases until it stabilizes at
\begin{equation}
\dot{I}_c^{\text{NESS}} = \frac{\sigma T_{\min}^\text{NESS}}{4\tau T}=\frac{\sqrt{1+\sigma}-1}{2\tau}.
\end{equation}
 In NESS, the fresh measurements are
no longer able to improve the quality of the estimate, i.e., to decrease the error variance $\langle [v(t)-\hat v(t)]^2 \rangle$ any further. Since $\dot W_{\max}=kT_{\min}\dot{I}_c$, the work-extraction rate also decreases until it stabilizes at
\begin{equation}
\dot W^\text{NESS} = kT_{\min}\dot{I}_c^{\text{NESS}} \frac{GR}{1+GR},
\label{eq:dotWNESS}
\end{equation}
see \refeq{eq:NESSwork}.

As in related studies \cite{horowitz+11,sagawa+12,bauer+12}, we can now define and study the \emph{information efficiency} $\eta$ of the demon,
\begin{equation}
\eta \equiv \frac{W}{kT I_c} \in [0,1].
\end{equation}
It measures the amount of extracted work per unit of received useful information.
An $\eta \approx 1$ means that the demon is close to saturating \refeq{eq:2ndlaw}, and is operating
at the limit of the generalized second law of thermodynamics. For our demons in NESS, we obtain the efficiency
\begin{equation}
\eta^\text{NESS} = \frac{T_{\min}}{T}\frac{GR}{1+GR},
\label{eq:etaNESS}
\end{equation}
using \refeq{eq:dotWNESS}. Hence, only a maximum work demon ($G \rightarrow \infty$) with $T_{\min}\approx T$ will operate at an information efficiency close to one. This corresponds to the poor measurement limit ($\sigma \approx 0$), and
a very small maximum work rate. A demon with access to almost perfect measurements ($\sigma\rightarrow \infty$) has $T_{\min}\approx 0$, and a very low information efficiency, $\eta \approx 0$.
Note also that a less aggressive demon (small $G$) has a lower efficiency, but that this is by choice: The transfer entropy rate is independent on $G$, and a smaller $G$ decreases $\dot W$, leading to a lower efficiency.
These observations are further elaborated upon and interpreted in Sections~\ref{sec:maxefficient} and
\ref{sec:impl}.

\section{Alternative switched control schemes}
\label{sec:maxefficient}
In this section, we consider three different switched work extraction schemes, to shed light on and to challenge the optimality of $W_{\max}$, and to establish some connections to previous work in the literature

\subsection{Maximizing information efficiency and the relation to Szilard's engine}
\label{sec:maxworkoverinfo}
As both the work extraction rate and transfer entropy rate are highest when the system is in equilibrium at temperature $T$, it may be tempting to run the feedback controller only when the system is close to equilibrium. Of course, as soon as the optimal feedback loop is closed, the effective temperature drops along the trajectory $T_{\min}$. But if
 the optimal feedback control is only applied for a very short time, say of duration $t_{\text{bur}} \rightarrow 0$, it holds
\begin{equation}
W_{\max}(t_{\text{bur}}) \approx k T I_c(t_{\text{bur}}) \approx \frac{\sigma k T}{4 \tau}t_{\text{bur}},
\end{equation}
since $T_{\min}(0)=T$. The work $\frac{\sigma k T}{4\tau}t_{\text{bur}}$ saturates the upper bound in \refeq{eq:mainres}, and has the largest possible information efficiency, $\eta\approx 1$. On the other hand, the amount of work is also very small since $t_{\text{bur}}$
is small. Nevertheless, if the system is allowed to relax back to thermal equilibrium again before the next feedback burst, it is possible to operate the feedback controller in a switched mode at the same efficiency as
feedback reversible discrete controllers \cite{horowitz+11,sagawa+12,szilard}, which saturate \refeq{eq:2ndlaw}.
To be specific: Let us apply $N$ bursts of feedback control, each burst of duration $t_{\text{bur}}$, and assume the time it takes for the system to relax back to thermal equilibrium is $t_{\text{rel}}$.
During the total time $N(t_{\text{bur}} + t_{\text{rel}})$ the amount of received useful information and extracted work become,
\begin{equation}
I_c  = NI_c(t_{\text{bur}}), \quad W \approx N kTI_c(t_{\text{bur}}),
\end{equation}
and $W/(kTI_c) = \eta \approx 1$. It should be noted that continuous extraction yields more work in the same amount of time,
$W_{\max}(N(t_{\text{bur}} + t_{\text{rel}})) > W$, albeit at a lower information efficiency. Nonetheless, it is possible to achieve a high efficiency also using continuous extraction, as already noted in Section~\ref{sec:infoflow}, but only in
the poor measurement limit, when the transfer entropy rate vanishes.

The switched controller discussed here mimics the discrete ones in that it tries to make a very short (sampled) measurement. It immediately acts on the obtained information and extracts the small work it can. If the controller waited to act on the system,
 its information would be less valuable due to the constant thermal fluctuations in $v$, as further discussed in the next subsection. Since the amount of extracted work is small, the system remains close to equilibrium, and
essentially the control is quasi-static. Therefore, the switched scheme here is analog to other maximum-efficiency
schemes which extract work quasi-statically, close to equilibrium, such as Szilard's engine \cite{szilard}.

\subsection{Collecting information before work extraction}
Another switched control strategy to consider starts with running the controller with $G=0$, just observing $v$ during some time $t_{\text{obs}}$. At time $t_{\text{obs}}$, the controller estimate $\hat v(t_{\text{obs}})$ has the estimation error variance $kT_{\min}(t_{\text{obs}})/C$. Since no control is applied, $T_C(t)=T$ throughout this mode and the capacitor remains in thermal equilibrium. When the estimate is considered good enough, one can
start the work extraction by switching on the control with $G\rightarrow \infty$. The hope with this scheme may be that by collecting more information before extraction, we may be able to increase the work amount.

As the extraction starts, the effective temperature $T_C$ will almost instantaneously drop from $T$ to $T_{\min}(t_{\text{obs}})$ while the work $\frac{1}{2}k(T-T_{\min}(t_{\text{obs}}))$ is retrieved.
If the controller is left in the mode $G\rightarrow \infty$ until some time $t\geq t_{\text{obs}}$,
the total amount of extracted work is
\begin{equation}
\int_{t_{\text{obs}}}^t \frac{k}{\tau} (T-T_{\min})\,dt' + \frac{1}{2}k(T-T_{\min}(t)).
\end{equation}
Compared to the maximum work $W_{\max}(t)$, it should be clear that an amount
$\int_0^{t_{\text{obs}}} \frac{k}{\tau} (T-T_{\min})\,dt'$  of work is lost by using this switched control strategy.
It does not pay off to wait and observe before applying the control.
The reason for the loss is that only a fraction of the received transfer entropy $I_c(t_{\text{obs}})$ can be used to extract work at time $t_{\text{obs}}$. A part of $I_c(t_{\text{obs}})$ is outdated when it comes to estimate the state $v(t_{\text{obs}})$, wherein the available energy is stored. In \cite{Mitter+05}, the information that is no longer useful for control is termed \emph{dissipated information}, and exact expressions for the amount is found there.

\subsection{Retrieving the non-equilibrium free energy}
As explained in \cite{esposito+11,horowitz+13}, a system's \emph{non-equilibrium free energy} can be used to upper bound the amount of work retrievable from a system in a particular (non-equilibrium) state. The non-equilibrium free energy \cite{Mitter+05,esposito+11}, is defined as $F\equiv U-TS$, where $U$ is the internal energy and $S$ the (Shannon) entropy. It is well known that $F$ attains a minimum, $F=F^\text{eq}$, in equilibrium. The amount of work attainable while forcing a system (in contact with a $T$-bath) to equilibrium $F^{\text{eq}}$
is bounded by the decrease of free energy $-\Delta F=F-F^\text{eq} = -\Delta U + T\Delta S$, see \cite{esposito+11,horowitz+13}.

After applying our maximum work extraction scheme for some time $t$, the system is clearly in a non-equilibrium state, since $\Delta U = \frac{k}{2}(T-T_{\min}(t))$ and $\Delta S = \frac{k}{2}\ln (T/T_{\min}(t))$. Hence, it would seem that in addition to the amount $W_{\max}(t)$, it should be possible to extract the work amount
\begin{equation}
-\Delta F = \frac{k}{2}(T_{\min}(t)-T) + \frac{kT}{2}\ln \left( \frac{T}{T_{\min}(t)}\right) \geq 0.
\end{equation}
Indeed, this is possible but only \emph{after} the allotted time interval $[0,t]$, and thus the amount $-\Delta F$
should not be counted in $W_{\max}(t)$.

To extract the work amount $-\Delta F$ without taking more measurements (without increasing $I_c$), we need to use a different type of work extraction than before. Inspired by \cite{delvenne+13}, we can apply a mechanical force and vary the plate distance of the capacitor. The work is then retrieved in two steps: First the plates are quickly (adiabatically) pulled apart until the effective temperature of the capacitor has increased to $T_C=T$. This requires the work $\Delta U$. Second, we let the plates slowly and isothermally (at temperature $T$) move together to their original position. This yields the work amount $T\Delta S$. Overall, the scheme yields the net work $-\Delta F$, without making any new measurements.

To summarize, it is possible to obtain the total work $W_{\max}(t)-\Delta F$ after $t$ time units of maximum work extraction and with the transfer entropy $I_c(t)$. Note, however, that for two reasons the additional work $-\Delta F$ does not fall within the scope of the problem we set out to study initially. First, the additional work cannot be retrieved using a (linear) current injection. This is understood since the expected voltage at time $t$ is zero, given the demon's knowledge, and nonlinear
 work extraction acting on the voltage variance is necessary.
 Second, the work is retrieved \emph{outside} of the allotted time interval $[0,t]$.
It should also be remembered that $-\Delta F$ is neglectable in comparison to $W_{\max}(t)$ for large $t$, since the latter grows linearly with $t$ and the former is bounded.
 Nevertheless, this example shows that an interesting problem for future research is to consider more general demons, with access to multiple actuation channels.

\section{The demon in non-equilibrium steady state and its physical implementation}
\label{sec:impl}
 The demon is traditionally seen as a little being observing the fluctuating system and acting on it so as to pump energy from it, apparently against the second law.
 In more modern treatments, it may be seen as a controller composed of a measurement device, a (digital or analog) computer that finds the most appropriate action, implemented by an actuator. The relevant information must be stored in the computer memory as long as it is deemed useful, and must be eventually discarded, or stored on an infinite memory tape.
 The Landauer-Penrose-Bennett treatment \cite{landauer61,bennett,penrose} of the demon considers the memory management combined with Landauer's principle, that erasing a bit of information in a memory at temperature $T$ must dissipate at least the work $kT \ln 2$ to the bath, as the key to reestablishing the second law.

\begin{figure}[tb]
\centering
  \psfrag{s}[][]{$\sigma$}
  \psfrag{Tmin}[][]{$\eta = T_{\min}/T$}
  \psfrag{dotI0}[][]{$\tau \dot I_c$}
  \psfrag{dotWmax}[][]{$\tau \dot W_{\max}/(kT)$}
  \includegraphics[width=1.00\hsize]{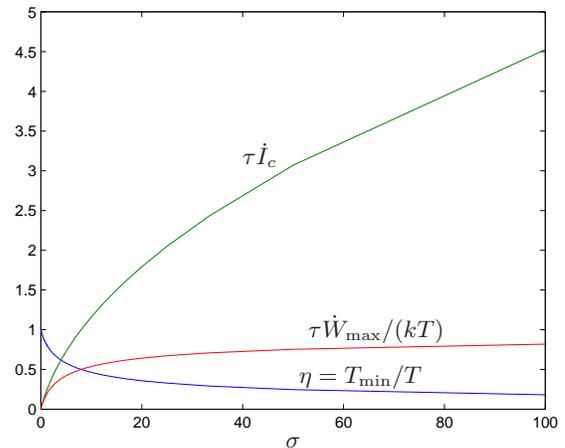}\\
  \caption{Information transmitted during a time constant of the system ($\tau\dot I_c$), normalized work extractable in the same time ($\tau\dot W_{\max}/(kT)$), and information efficiency ($\eta$), for a maximum work demon at NESS, as function of measurement quality $\sigma$. We note that even a demon with large amounts of information cannot extract more work than $kT$ during a time constant of the system, in accordance with \refeq{eq:Wmaxupper}.}\label{fig:info_vs_sigma}
\end{figure}

 In the spirit of \cite{jarzynski+12,jarzynski+13,esposito+13,horowitz+13}, we next find explicit physical devices implementing the demon, which allows for a detailed discussion of the energy or entropy flows inside the demon. To simplify the presentation, we only consider the NESS, and avoid writing out the superscript `NESS' on the quantities in this section. The key quantities of interest in NESS, derived in the previous sections and which our implementation of the demon will realize, are
\begin{align}
\dot I_c & = \frac{\sqrt{1+\sigma}-1}{2\tau} \\
T_{\min} & =  \frac{2T}{\sqrt{1+\sigma}+1} \\
\dot W & = kT_{\min}\dot I_c\frac{GR}{1+GR}  \leq \dot W_{\max} = kT_{\min}\dot I_c.
\end{align}
In Fig.~\ref{fig:info_vs_sigma}, normalized versions of these quantities are plotted as a function of measurement quality $\sigma$.
In particular, we emphasize that an infinite transfer entropy rate
(perfect measurements) \emph{does not} result in an infinite work rate. The intuitive explanation is that the demon acts on the capacitor, which in equilibrium has the energy $\frac{1}{2}kT$ and equilibrate with the bath with the time constant $\tau$. A very large information supply will allow the demon to extract almost all of the energy $\frac{1}{2}kT$, but it will nevertheless need to wait a time of order $\tau$ before the bath has refilled the capacitor with new energy to be extracted. The fundamental upper bound on the work rate is therefore proportional to $1/\tau$,
\begin{equation}
\dot W_{\max}=kT_{\min}\dot I_c = \frac{k}{\tau}(T-T_{\min})\leq \frac{kT}{\tau},
\label{eq:Wmaxupper}
\end{equation}
which is reached with equality by a maximum work demon with infinite transfer entropy rate, i.e., $G\rightarrow\infty$ and $\sigma\rightarrow \infty$. Such a demon comes at a very high implementation cost since it needs to process a huge amount of information, as will be further discussed below. As was already found in Section~\ref{sec:infoflow}, we also see that in the poor measurement limit ($\sigma\approx 0$), the capacitor is close to equilibrium with the bath, $T_{\min}\approx T$, even under maximum work extraction. This makes the information efficiency as large as possible, $\eta = W/(kTI_c)\approx 1$, although the extracted power is small. This observation is in accordance with Section~\ref{sec:maxworkoverinfo}, where we also saturated the information efficiency by only processing small amounts of information and using it to the maximum.

Kalman-Bucy demons, as we call those demons that implement a Kalman-Bucy filter together with a linear control law (\refeq{eq:feedback}), turn out to be implemented with simple linear circuit elements. As such they include a real (resistive) impedance $Z$ that is accountable for the energy  absorbed by the demon, and is associated with random Johnson-Nyquist thermal noise, following fluctuation-dissipation theorem. We naturally assume that this thermal noise is precisely responsible for the measurement noise $\sqrt{V_\text{meas}}w_\text{meas}$. One may therefore define an effective demon temperature $T_\text{dem}$ that suitably explains the observed level of measurement noise as
\begin{equation}
T_\text{dem}\equiv \frac{V_\text{meas}}{2kZ}.
\label{eq:Tdem}
\end{equation}
We next turn to analyzing the electrical circuit used to realize the above expressions.

\begin{figure}[tb]
\centering
  \includegraphics[width=1.00\hsize]{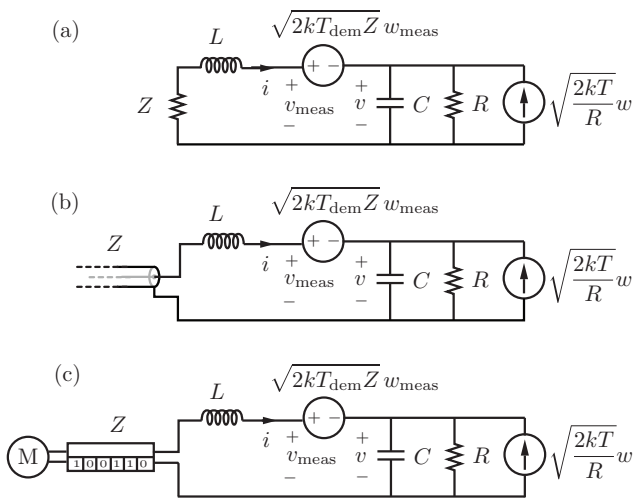}\\
  \caption{Three exact electric implementations of the Kalman-Bucy demon, with different realizations of the purely resistive element   $Z$. The current $i$ through the inductive element is proportional to the optimal (Kalman-Bucy) estimate $\hat v$, and
  $Z$ creates measurement noise. To realize $Z$, implementation (a) uses a normal resistor, (b) uses a lossless transmission line as memory element and energy storage, and (c) uses an abstract computer with memory element and a motor M. In the three cases, $Z$ is perceived by the circuit as a resistance $Z$ of temperature $T_\text{dem}$.}\label{fig:RLCcircuit}
\end{figure}

\subsection{Implementation of the demon}
\label{sec:demoneff}
To exactly implement the Kalman-Bucy filter and the control law (\refeq{eq:feedback}), only a resistive and an inductive element are needed, as illustrated in Fig.~\ref{fig:RLCcircuit}. The inductance $L$ and the resistance $Z$ should be chosen as
\begin{equation}
\begin{aligned}
L  & =  \frac{CZ_{\min}}{G} \\
Z  & =  \frac{1}{G} \left(1 + \frac{Z_{\min}}{R}\right) + Z_{\min} \geq Z_{\min},
\end{aligned}
\label{eq:opt_circuit}
\end{equation}
where
\begin{equation}
Z_{\min}= \frac{R}{\sqrt{1+\sigma}-1}>0
\end{equation}
is the resistance of a demon extracting work at the maximum possible rate (maximum power, at $G \rightarrow \infty$).
Hence, a Kalman-Bucy demon that would like to implement a less
aggressive control law (finite $G$) should use a resistance \emph{larger} than $Z_{\min}$.
One may interpret the inductor $L$ as both actuator and dynamic state of the Kalman-Bucy demon, since the best instantaneous estimate $\hat v(t)$ of the capacitor voltage is proportional to the current $i(t)=-G\hat v(t)$. The demon temperature can now be explicitly computed as
\begin{equation}
T_\text{dem} = \frac{T_{\min}}{2} \frac{GR}{GR+\sqrt{1+\sigma}} < T.
\label{eq:TD}
\end{equation}
We note in particular that there is always a temperature gradient from the demon to the bath, driving the energy flow $\dot W$.

Let us insist that the element $Z$, although required to behave externally as a resistance of temperature $T_\text{dem}$, is otherwise unconstrained. It may be a simple resistor, which burns the received energy flow $\dot W$ into heat. It may also be realized as a semi-infinite lossless transmission line of characteristic impedance $Z$, in which injected energy is stored and disposed of as a travelling wave satisfying the telegrapher's equation. A transmission line, a well-known lossless realization of a resistance~\cite{cheng,Sandberg+11}, is the electric equivalent of an elastic string, itself a well-known lossless model of friction \cite{Lamb00}. Assuming a wave propagation velocity of
$c$ means that the control actuation applied and the power extracted $\Delta t$ time units ago is stored in the
line voltage $Zi(t-\Delta t)$ and line current $i(t-\Delta t)$, to be found a distance $c \Delta t$ down from the terminal connected to $L$. In this sense, we can think of the lossless line as a model of a continuous and infinite memory tape, where all past demon control actions and energy are stored.
Finally, $Z$ can be a `load' resistance, i.e., a machine, comprising for instance a computational device, memory, a motor, etc. that converts at least part of the electric energy $W$ into another useful form, such as the kinetic energy of a wheel maintained in motion by the motor against adversary forces.  We do not here go into the details of how to actually build such a device, albeit in any case it should externally look like a resistance $Z$.
A simple example of an electro-mechanical device converting electrical energy into work while externally looking like a resistor is given in \cite{delvenne+13}.
All of these different interpretations of $Z$ are illustrated in Fig.~\ref{fig:RLCcircuit}.

\subsection{Implementation costs of the demon}
\label{sec:TDimpl}
From \refeq{eq:TD}, it is seen that as $G$ increases, the Kalman-Bucy demon temperature $T_\text{dem}$  increases and at maximum power extraction, $T_\text{dem}$ is half the effective capacitor temperature, i.e.,
\begin{equation}
T_\text{dem}=\frac{1}{2}T_C = \frac{1}{2}T_{\min} =  \frac{T}{\sqrt{1 + \sigma}+1}.
\label{eq:TDTmin}
\end{equation}
To apply an external potential (the demon temperature) that is half of the source potential (the capacitor temperature) to generate maximum power transfer has also been observed in previous work in other settings, see \cite{vandenbroeck05,esposito+09}.
The maximum power demon is electrically equivalent to just a resistance $Z_{\min}$ in series with a vanishing inductance, $L \rightarrow 0$. While the inductance vanishes, the current through it increases as the demon becomes more aggressive ($G \rightarrow \infty$), resulting in a non-vanishing stored energy $L\langle i^2\rangle=kT_\text{dem}$. One cannot simply remove the inductance from the circuit in general, for a correct representation. However, in NESS the inductance plays no role in the energy and entropy flows. To simplify the presentation, we will only consider this maximum power case ($Z=Z_\text{min}$, $\dot W=\dot W_\text{max}$), since finite $G$ can be treated analogously.

We first assume that $Z$ is a pure resistor, the simplest situation to analyse. If the demon has access to a bath $T_\text{dem}$, then the resistor creates a noise of same temperature and burns  $\dot W$ into heat as required. If the demon has only access to the same bath as $R$, of temperature $T$, then one may refrigerate the resistor down to $T_\text{dem}$, and evacuate the heat $\dot W$ from $T_\text{dem}$ to $T$. This
can be done by a Carnot ideal refrigerator with an external power supply of
\begin{equation}
\left(\frac{T}{T_\text{dem}}-1\right)\dot W = \left(\frac{2}{\eta}-1\right)\dot W.
\end{equation}
This extra supply is the cost of maintenance of the NESS, showing clearly the demon, in this implementation is overall active unless it can rely on a colder heat bath than $T$. In the poor measurements limit ($\sigma \approx 0$, $T \approx T_{\min} = 2T_\text{dem}$), the refrigeration cost approaches $\dot W$. With good measurements, the refrigeration cost is much larger than $\dot W$. A maximum power demon with a high-quality measurement device has a small information efficiency $\eta$, and requires a large external power supply. In all cases, the \emph{total energy} dumped to the $T$-bath takes the simple form
\begin{equation}
\dot W_\text{impl} = \left(\frac{T}{T_\text{dem}}-1\right)\dot W +\dot W = 2kT \dot I_c,
\label{eq:dumped}
\end{equation}
since $\dot W = kT_{\min}\dot I_c$ and $2T_\text{dem}=T_{\min}$. This is a measure of the implementation cost of the demon, as further discussed below.

Let us now consider the model of the lossless transmission line of characteristic impedance $Z$. Let us recall that a transmission line is akin to an ideal memory tape, with information encoding past measurements of the system travelling away from the circuit towards infinity, and random white noise travelling from infinity into the circuit. A crucial observation is that the useful information retrieved from the measurement, the transfer entropy $I_c$, is embedded into a signal of larger information rate, and it is that total signal that the memory tape has to manage. We show in Appendix~\ref{sec:inforate} from a spectral analysis of the signal into the tape, that the total information rate entering the memory is
\begin{equation}
\textrm{InfoRate}=\int_0^{\infty} \ln\left(1+\frac{\dot{W}(f)}{kT_\text{dem}}\right) df,
\label{eq:bits}
\end{equation}
where (with some abuse of notation) $\dot{W}(f)$ is the power spectral density of $\dot W$
at the Fourier mode of frequency $f$, so that $\dot W=\int \dot{W}(f) df$. From the inequality $\ln(1+x) \leq x$, we easily see
\begin{equation}
\textrm{InfoRate}\leq \frac{\dot W}{kT_\text{dem}} = 2\dot I_c.
\label{eq:bits2}
\end{equation}

Let us now assume a finite-length memory tape, requiring to erase bits at the same rate they enter the line. This can be done in a simple way by ending the finite-length line with a resistance of exact value $Z$, burning all energy and information on the line. We here recover the simple resistor analysis, where the line only creates a delay in the dissipation. We may imagine many other ways to erase information on the line, but any such mechanism should respect Landauer's principle \cite{landauer61} and dispose \emph{at least $kT$ times InfoRate} to the bath of temperature $T$. Therefore, using \refeq{eq:bits2},
\begin{equation}
\textrm{LandauerCost} \leq \frac{T}{T_\text{dem}} \dot W = 2kT\dot I_c.
\label{eq:Landauer}
\end{equation}
Landauer's cost therefore coincides with the implementation cost (\ref{eq:dumped}) derived above in the poor measurements limit ($T\approx T_{\min}=2T_\text{dem}$, $\sigma \approx 0$) where InfoRate$\approx \frac{\dot W}{kT_\text{dem}}=2\dot I_c$. In this case, every bit of transfer entropy measured from the system allows to retrieve an energy $kT \ln 2$ from the system, but comes with another bit of `useless' information to be stored, and eventually erased, by the memory. Therefore the demon earns $kT \ln 2$ but has to dispose at least $2kT \ln2$ to the heat bath. The second law is therefore true with a wide margin.

Further away from equilibrium ($\sigma>0$), the strict inequality in Eq.~(\ref{eq:Landauer}) seems to suggest that Landauer's bound leaves potential room for a more clever information-erasure mechanism than pure refrigeration. Nevertheless, we show below that (\ref{eq:dumped}) provides the correct minimum implementation cost. The strict inequality thus arises from the conservativeness of Landauer's bound far from equilibrium, when the information flow is high and fast erasure is required. In Appendix~\ref{sec:inforate}, we postulate a non-equilibrium extension of Landauer's principle, demanding proportionally more power consumption for high rate erasure, which allows us to recover the same implementation cost as for the resistor. Even if the demon chooses to cover parts of $\dot W_\text{impl}$
with the energy flow $\dot W$, gained from the system, an extra $\dot W_\text{impl}-\dot W = (2T-T_{\min})k\dot I_c> \dot W$ has to be supplied by an external source of power.

We next turn to an interpretation where $Z$ is any device that attempt to convert $\dot W$ to mechanical work. In the absence of concrete implementation details, we can only proceed
from general thermodynamic principles. No matter how it is actually built, the element externally should appear as a heat bath of temperature $T_\text{dem}$. Hence, since energy at a rate $\dot W$ disappears into $Z$, an entropy at a rate
\begin{equation}
\dot S = \frac{\dot W}{T_\text{dem}}
\end{equation}
is at least being created inside $Z$. Now, we assume that $Z$ is composed of a steady state finite device in communication with a bath of temperature $T$. According to the second law, this can only be achieved if we dispose at least an amount
\begin{equation}
\dot W_\text{impl}= \frac{T}{T_\text{dem}}\dot W = 2kT\dot I_c
\end{equation}
 of energy into the heat bath of temperature $T$, a general implementation cost of the demon, whether it consists of a refrigerated resistor, a continuous memory tape with erasure mechanism, or any other implementation.

 In conclusion, any maximum power demon for a linear system must be active, with at least an extra power supply $\dot W_\text{impl} - \dot W =(2T-T_{\min})k\dot I_c > \dot W$. The second law is thus safe with a margin.
 As a final remark, we note that less aggressive Kalman-Bucy demons ($G < \infty$) have a lower effective temperature, see \refeq{eq:TD}, and lower information efficiency, see \refeq{eq:etaNESS}, and thus spend even more power into heat, $\frac{T}{T_\text{dem}}\dot W > 2kT \dot I_c$. This has the surprising consequence that the maximum power demon is the cheapest of all Kalman-Bucy demons, cheaper in particular than the pure observer ($G \rightarrow 0$, no control).

\section{Discussion}
\label{sec:disc}
 The first finding of our study is that the maximum work a demon can extract from a thermally fluctuating system over a time interval $[0,t]$ takes the form $W_{\max}(t)=\int_0^t kT_{\min}\dot I_c\,dt'$. Here $T_{\min}<T$ for $t'>0$. One may have expected $T_{\min}=T$ for all $t'$, i.e., that every unit of transfer entropy corresponds to $kT$  units of potential work, as suggested by previous works \cite{sagawa+12}. But if time is limited, there is a diminishing return on the received information. In fact, $W_{\max}(t)\leq kTt/\tau$, where $\tau$ is the time constant of the fluctuating system. This bound becomes tight if the demon has access to very accurate measurements, so that the transfer entropy rate $\dot I_c$ tends to infinity. The work $W_{\max}$ remains finite since $T_{\min}$ tends to zero. Using optimal filtering theory, we show that $T_{\min}$ can be interpreted as an out-of-equilibrium effective temperature of the maximally cooled fluctuating system. An intuitive explanation as to why a very well informed demon cannot extract unbounded amounts of energy in finite time, even with access to unbounded amounts of information, is that the fluctuating system is of low-pass character. That is, all thermal fluctuations of frequency above $1/\tau$ are effectively attenuated by the system itself and the corresponding (unbounded) energy is kept beyond reach of the demon. We recover a work of $kT$ per unit of useful information every time that the demon extracts work at an infinitely slow rate, either by intermittent, bursty control,  a continuous strategy that only uses a fraction of retrieved information, or because the measurement is of poor quality. We underline that our main conclusions apply, not only to the capacitance obeying a linear scalar Langevin equation that served as motivating example, but more generally to any system undergoing small thermal fluctuation around a minimum energy level.

Our second contribution is to use control theory to characterize and interpret the feedback protocol the demon should apply to reach the upper limit $W_{\max}$. The protocol is a linear feedback law based on the Kalman-Bucy estimate of the system state. The so-called separation principle shows that the demon should use all the received information to first optimally estimate the current state of the system, and then quickly pull out the energy it can. Since our system is subjected to continuous thermal fluctuations, old information is less useful and there is no reason to wait before exploiting it. We also propose an easily analyzed family of demons, the Kalman-Bucy demons. These demons interpolate with less and less aggressive control strategies between the maximum power demon and a demon that only observes the system without acting on it. The family illustrates the trade-off obtained when we give up maximum power extraction and remain close to equilibrium, and allows for a better comparison with the literature on many aspects. We believe that Kalman-Bucy demons and variants can play a role similar to the Szilard's engine and its variants in illustrating and understanding the fundamental interactions between thermodynamics and information theory, in particular in non-equilibrium situations.

 Kalman-Bucy demons offer simple physical implementations.
 We find that any such implementation is necessarily active, i.e., consumes power from an external source
that exceeds the power retrieved from the system. This makes Kalman-Bucy demons necessarily energy-deficient, in whatever implementation. We can understand this since the demons are out of equilibrium with the system even when the system is close to equilibrium with the bath, and we need to spend work to keep them in such states.
More specifically, the extra cost can be interpreted as the cost of a noise suppression mechanism, e.g., through refrigeration or memory erasure, and the deficit can be interpreted as the fact that only part of the information that must be handled by the demon's memory (InfoRate) is useful to retrieve energy ($\dot I_c$).

Finally, we believe that our results exemplify how key tools from control theory, such as (continuous-time) Kalman-Bucy filtering, the separation principle, and circuit realization, can contribute to stochastic thermodynamics and statistical mechanics.

\emph{Acknowledgements.}--- The authors would like to thank Jordan Horowitz for many helpful discussions and suggestions.
H.S.\ is supported by the Swedish Research Council under grants 2009-4565 and 2013-5523. J.-C.D.\ is supported by the Interuniversity Attraction Pole `Dynamical Systems, Control and Optimization (DYSCO)', initiated by the Belgian State, Prime Minister's Office. S.K.M.\ is supported in part by Siemens Corporate Research Grant, `Methods for Optimal Control in Grids with Storage', and 
NSF Grant EECS-1135843, `Smart Power Systems of the Future: Foundations for Understanding Volatility and Improving Operational Reliability'.

\hrulefill

\appendix

\section{The transfer entropy $I_c(t)$}
\label{sec:TE}

In order to motivate our definition of $I_c(t)$ for the continuous-time system in~\refeq{eq:RCcircuit},
we first consider the notion of transfer entropy in the more familiar context of a
controlled \emph{discrete-time} system with signal and observation processes subject to
additive noise.  In this, the signal and observation sequences, $(X_k; k=0,1,\ldots)$
and $(Y_k;k=0,1,\ldots)$, satisfy the following equations:
\begin{equation}
\begin{aligned}
X_0     & = W_0 \\
X_{k+1} & = f_k(X_0^k) + u_k(Y_0^k) + W_{k+1}  \\
Y_k     & =  g_k(X_0^k) + Z_k,
\end{aligned}
\label{eq:discsys}
\end{equation}
where $(W_k; k=0,1,\ldots)$ and $(Z_k;k=0,1,\ldots)$ are independent white noise
sequences.  (By this, we mean that the random variables $\{W_k,Z_l, 0\le k,l <\infty\}$
are independent.)  We assume that the functions $f_k$, $g_k$ and $u_k$, and the
distributions of the noise sequences are such that, for some $0\le N < \infty$, the
mutual information between $X_0^N$ and $Y_0^N$ is finite:  $I(X_0^N;Y_0^N)<\infty$.

$I(X_0^N;Y_0^N)$ has its origins in two components of entropy exchange between $X$ and
$Y$: one from $X$ to $Y$ through the observation function $g_k$, the other from $Y$ to
$X$ through the control function $u_k$.  The first of these is called in
\cite{sagawa+12} the \emph{transfer entropy}, and is the component most important to us
here since it determines the information about the statistical mechanical system modelled
by $X$ made available to the demon by the partial observations $Y$. It is sometimes
called the \emph{directed information}. In the context of (\ref{eq:discsys}) the
transfer entropy is defined in the following way \cite{sagawa+12}:
\begin{equation}
I_c(n)\equiv \sum_{k=0}^n I(X_0^k;Y_k|Y_0^{k-1}), \quad\text{where }Y_0^{-1}\equiv0.
\label{eq:Icdisc}
\end{equation}

\begin{prop} \label{prop:snform}
For the system of (\ref{eq:discsys}) and any $0\le n\le N$,
\begin{equation}
I_c(n)=I(W_0^n;Y_0^n).
\label{eq:snform}
\end{equation}
\end{prop}

\noindent{\bf Proof} (Induction) The case $n=0$ is trivial.  Suppose then that
(\ref{eq:snform}) is true for some $0\le n <N$.  It follows from two applications of
the chain rule of mutual information (see, for example, Theorem~2.5.2 in \cite{coverthomas}) that
\begin{equation}
\begin{aligned}
I(W_0^{n+1};Y_0^{n+1})
 & = I(W_0^{n+1};Y_0^n) + I(W_0^{n+1};Y_{n+1}|Y_0^n) \\
 & = I(W_0^n;Y_0^n) + I(W_{n+1};Y_0^n|W_0^n) \\
 & \qquad + I(W_0^{n+1};Y_{n+1}|Y_0^n) \\
 & = I(W_0^n;Y_0^n) + I(W_0^{n+1};Y_{n+1}|Y_0^n),
\end{aligned}
\end{equation}
the last step resulting from the independence of $W_{n+1}$ and $(W_0^n,Y_0^n)$.  It
thus remains to prove that $I(W_0^{n+1};Y_{n+1}|Y_0^n)=I(X_0^{n+1};Y_{n+1}|Y_0^n)$,
but this follows from ``sufficient statistics'' arguments based on the fact that there
exist maps $F_{n+1}$ and $G_{n+1}$ such that
\begin{equation}
\begin{aligned}
X_0^{n+1} & = F_{n+1}(W_0^{n+1},Y_0^n) \quad
  \text{and}\\
  W_0^{n+1} & = G_{n+1}(X_0^{n+1},Y_0^n).
\end{aligned}
\end{equation}

Proposition~\ref{prop:snform} shows that the transfer entropy is equal to the mutual
information between the observation $Y$ and the signal noise $W$.  (We regard the
signal initial condition $X_0$ as being part of this noise sequence.)
Unlike the signal itself, $W$ is not affected by the action of the control term
$u_k(Y_0^k)$ in (\ref{eq:discsys}), and so is not influenced by the second component of
entropy exchange identified above.  We use Proposition \ref{prop:snform} to motivate
our definition of transfer entropy for the continuous-time system in Eq.~(\ref{eq:RCcircuit}).

Although it can be formally defined through its (constant) power spectral density,
continuous-time Gaussian white noise does not have sample paths with any reasonable
properties.  The values it takes at two distinct times (no matter how close) are
independent, ``infinite-variance'' random variables.  To give precise meaning to
equations such as (\ref{eq:RCcircuit}) we need to use stochastic calculus. This expresses
both equations in (\ref{eq:RCcircuit}) as \emph{integral} equations, thereby replacing
the white noise processes $w$ and $w_\text{meas}$ by Brownian motion processes $B$ and
$B_\text{meas}$.  (In a formal sense, $w=dB/dt$ and $w_\text{meas}=dB_\text{meas}/dt$,
although neither $B$ nor $B_\text{meas}$ is actually differentiable.)  The measurement
voltage, $v_\text{meas}$, is replaced by its integral form, which we denote $Y$.
Eq.~(\ref{eq:RCcircuit}) is thereby replaced by the following pair of equations:
\begin{equation}
\begin{aligned}
\tau v(t) & = \tau v(0) +\int_0^t (Ri-v)\,dt' + \sqrt{2kTR}B(t)  \\
Y(t)      & = \frac{1}{\sqrt{V_\text{meas}}}\int_0^t vdt' + B_\text{meas}(t).
\end{aligned}
\label{eq:intsys}
\end{equation}

We define the mutual information between $v$ and $v_\text{meas}$ to be that between $v$ and $Y$.
As above, the latter has its origins in two exchanges of entropy: one from $v$ to $Y$
through the observation mechanism, the other from $Y$ to $v$ through the control; only
the first of these, the transfer entropy, is relevant to the demon. Motivated by
Eq.~(\ref{eq:snform}), we define the transfer entropy of the system (\ref{eq:intsys}) as
follows:
\begin{equation}
I_c(t) \equiv  I((v(0),(B)_0^t);(Y)_0^t).  \label{eq:CTdef}
\end{equation}
This can be found by a classical result dating back to \cite{duncan70}, which
appears in a fairly general form in \cite{liptser+2001}. It is expressed there in the
context of a problem of communication across a channel subject to additive Gaussian white
noise.  A ``message'' signal $\theta(t)$ is encoded, by a mechanism that has access to
the output of the channel $\xi(t)$, to produce a channel input signal, $a_t(\theta,\xi)$.
This has finite variance and is \emph{non-anticipative} in the sense that, for each time
$t$, $a_t(\theta,\xi)$ depends only the past and present of $\theta$ and $\xi$
($(\theta)_0^t$ and $(\xi)_0^t$).  Theorem~16.3 in \cite{liptser+2001} derives an
explicit form for the mutual information between the process segments $(\theta)_0^t$ and
$(\xi)_0^t$; in fact
\begin{equation}
I((\theta)_0^t;(\xi)_0^t)
  = \frac{1}{2} \int_0^t \langle [a_{t'}(\theta,\xi) - \hat a_{t'}(\xi)]^2\rangle\, dt',
\label{eq:LStransfer}
\end{equation}
where $\hat a_{t}(\xi)$ is the $(\xi)_0^t$-conditional mean of $a_{t}(\theta,\xi)$.

In the context of Eqs.~(\ref{eq:intsys}) and (\ref{eq:CTdef}), the message signal is the
pair $(v(0),B)$, the input to the channel is $v/\sqrt{V_\text{meas}}$, and the output of
the channel is $Y$.  The representation $v(t)=\sqrt{V_\text{meas}}\,a_t((v(0),B),Y)$
is made explicit by the first equation in (\ref{eq:intsys}).  The non-anticipative
condition of Theorem~16.3 in \cite{liptser+2001} is satisfied if the injected current
$i$ is itself non-anticipative (i.e.~if $i(t)$ depends only on the past and present of
$Y$).  This is a natural condition to impose on the demon---it should use only past and
present measurements of $Y$ when deciding what current to inject at time $t$.  This
condition is certainly satisfied by the feedback control of Eq.~(\ref{eq:feedback}).
Substituting these terms into Eqs.~(\ref{eq:CTdef}) and (\ref{eq:LStransfer}), we obtain the
following explicit form for the continuous-time transfer entropy
\begin{equation}
\begin{aligned}
I_c(t)
  &  = \frac{1}{2V_\text{meas}}\int_0^t \langle [v-\hat{v}]^2 \rangle dt' \\
  &  = \frac{1}{2V_\text{meas}}\int_0^t \frac{kT_\text{min}}{C} dt' \\
  &  = \frac{\sigma}{4 \tau} \int_0^t \frac{T_{\min}}{T}\,dt'.
\end{aligned}
\label{eq:CTtransfer}
\end{equation}
This is not dependent on the value of the control gain $G$.  In fact it would take the
same value with any control regime for which the resulting process $v$ satisfied the
finite variance and non-anticipative conditions.  This is true, for example, for a
large class of nonlinear feedback controls.

\section{The Hamiltonian case in higher dimension}
\label{sec:multidim}
Let us consider the more general case where the capacitor in~\refeq{eq:RCcircuit} is replaced by a Hamiltonian system.
We assume a quadratic Hamiltonian, $H(x) = \frac{1}{2}x^T K x$, where $x^T = [q^T \quad p^T]\in\mathbb{R}^{2n}$
is a point in the phase space with generalized positions $q$ and momenta $p$, and $K \in\mathbb{R}^{2n\times 2n}$
is a symmetric positive-definite matrix. Hamilton's equations under the influence of a generalized external force $B_u u(t)$ (the constant matrix $B_u\in\mathbb{R}^{2n}$ determines which coordinates are directly affected), applied by the demon, now reads
\begin{equation}
\begin{aligned}
\dot x & = J\nabla H(x) + B_u u \\
y & = B_u^T \nabla H(x),
\end{aligned}
\end{equation}
where $J = -J^T = \bigl[ \begin{smallmatrix}0 & I_{n} \\ -I_{n} & 0\end{smallmatrix} \bigr]$, and $y$ is the generalized velocity conjugate to $u$. That is, $\dot H(t) = y(t)u(t)$
is the rate of work applied to the system. Now, $\nabla H(x) = Kx$, and the Hamiltonian system is a \emph{linear dynamical system}.

We connect the Hamiltionian system to a heat bath of temperature $T$ and with viscous friction coefficient
$r>0$ producing a dissipative force in the
direction $B \in \mathbb{R}^{2n}$. We obtain \cite{delvenne+13}
\begin{equation}
\begin{aligned}
\dot x & = (J - D) Kx + B_u u + B \sqrt{2k T r}w, \\
\langle x(0)\rangle & = 0, \quad  \langle x(0)x(0)^T \rangle = kTK^{-1}, \\
y & = B_u^T K x, \\
y_\text{meas} & = B^T K x + \sqrt{V_\text{meas}}w_\text{meas},
\end{aligned}
\label{eq:noisyHamilton}
\end{equation}
where $x(0)$ is Gaussian, $w$ and $w_\text{meas}$ uncorrelated Gaussian white noise, $D = r BB^T$ is the dissipation and $B \sqrt{2k T r}w$ models the corresponding thermal fluctuation. We have also assumed a scalar noisy measurement $y_\text{meas}$ of the generalized velocity conjugate to the dissipative force \footnote{The dissipative force does not need to be parallel with the actuation force. This was the case for the overdamped Langevin equation but is not necessary in higher dimension.},
which is available to the demon. In the following, it is assumed the system in \refeq{eq:noisyHamilton}
 is \emph{controllable} and \emph{observable} \cite{astrom+08}. That is, in the absence of noise ($w=w_\text{meas}=0$),
it is possible to force the system to $x=0$ in arbitrarily short time from any initial point using some force $u$, and it is possible to determine $x(t)$ exactly given an arbitrarily short measurement trajectory $(y_\text{meas})_{t-\epsilon}^{t+\epsilon}$, $\epsilon>0$. If these assumptions do not hold, it means that there are system coordinates that are either invisible to, or beyond the influence of, the demon. Such degrees of freedom can systematically be eliminated to create a \emph{minimal model}, see, for example, \cite{astrom+08}.

Let us denote the second moment of the phase space coordinate by $X(t)\equiv \langle x(t)x(t)^T \rangle \in \mathbb{R}^{2n \times 2n}$. Then the internal energy can be written as
$U(t)=\langle H(t)\rangle = \frac{1}{2} \text{Tr} (KX(t))$. The first law of thermodynamics reads \cite{delvenne+13}
\begin{equation}
\begin{aligned}
\dot U & = \dot Q - \dot W \\
\dot Q & = k T  \text{Tr} (KD) - \text{Tr} (K D K X) \\
\dot W & = -\langle uy \rangle,
\end{aligned}
\label{eq:1stlaw}
\end{equation}
where $\dot Q$ is the expected energy exchange rate with the heat bath, and  $\dot W$ is the expected work extraction rate.
We note that in thermal equilibrium ($\dot Q=\dot W = 0$) we have $X = k T K^{-1}$, and the internal energy
is $U = n k T$, in accordance with the equipartition theorem. We say the internal energy is \emph{equipartitioned}
when $X$ takes the form $k T K^{-1}$ for some scalar temperature $T$.

Similarly to the scalar case, we can determine the smallest achievable second-moment of the phase space coordinate,
$X_{\min}$, under all possible causal feedback laws $u(t)=f_t(y_\text{meas})$. It satisfies the filter Riccati equation
\begin{equation}
\begin{aligned}
\dot X_{\min} & =  (J- D) KX_{\min} + X_{\min} K (J- D)^T \\ & \quad + 2k  T D - X_{\min} KB V_\text{meas}^{-1}B^TK X_{\min}, \\ X_{\min}(0)& = X(0) = k T K^{-1}.
\end{aligned}
\label{eq:multiricc}
\end{equation}
As before, the internal energy for the controlled system must obey a bound, $U(t) \geq U_{\min}(t) \equiv \frac{1}{2}\text{Tr}(KX_{\min}(t))$.
The assumption on controllability and observability ensures that
there exists a feedback control that drives the internal energy to the limit $U(t)=U_{\min}(t)$.
Just as in the scalar case, one such
control is a high-gain feedback from the Kalman-Bucy state estimate $\hat x$. For example, one can use $u(t)=-B_u^T G \hat x(t)$, for a suitably chosen large positive-definite gain matrix $G$.

Using the first law of thermodynamics, \refeq{eq:1stlaw}, we can quantify the maximum possible amount of extractable work by
\begin{equation}
\begin{aligned}
W_{\max}(t) & = \int_0^t -\dot U_{\min} + k T \text{Tr} (KD) \\
& \qquad  - \text{Tr} (K D K X_{\min})\,dt' \\
& = \frac{1}{2} \int_0^t \text{Tr} (KX_{\min} KB V_\text{meas}^{-1}B^TK X_{\min})\,dt'.
\end{aligned}
\label{eq:multimaxwork}
\end{equation}
The transfer entropy from $x$ to $y_\text{meas}$ in \refeq{eq:noisyHamilton} (analogously to
 Eq.~\eqref{eq:mutualinfo}) is
\begin{equation}
I_c(t)=\frac{1}{2}\int_0^t\text{Tr} (KB V_\text{meas}^{-1}B^T K X_{\min})\,dt',
\end{equation}
which clearly has many factors in common with $W_{\max}$.
Nevertheless, in the matrix case, the integrand in $W_{\max}$ does not generically factorize into a product of the transfer entropy rate and a scalar temperature, unless $X_{\min}$ is equipartitioned, $X_{\min}=kT_{\min}K^{-1}$ for some
scalar $T_{\min}$. However, it is possible to define a useful scalar instantaneous
\emph{effective} temperature for arbitrary $X$ as follows. By assuming $\dot Q = 0$ instantaneously in
\refeq{eq:1stlaw}, we \emph{define} the effective temperature in the state $X(t)$ as
\begin{equation}
T_X(t) \equiv \frac{\text{Tr}[KDKX(t)]}{k\text{Tr}(KD)}.
\end{equation}
The physical intuition behind the definition is that if the system has covariance $X(t)$ and
is connected to a heat bath of temperature $T_X(t)$, along the direction $B$, then there is no instantaneous heat exchange between the system and the heat bath. This effective temperature does not depend on the friction coefficient $r$, and transforms  \refeq{eq:1stlaw} into a Fourier-like heat conduction equation as in the scalar case (see \refeq{eq:heatflows}):
\begin{equation}
\dot Q  = k \text{Tr} (KD) (T- T_X).
\label{eq:fourier}
\end{equation}
If the system is equipartitioned at temperature $T$, then $T_X= T$.

Using the effective temperature and applying the Cauchy-Schwarz inequality ($\text{Tr}(AB)^2 \leq \text{Tr}(AA^T)\text{Tr}(BB^T)$) we obtain the general lower bound,
\begin{equation}
k\int_0^t T_{\min} \dot{I}_c\, dt' \leq W_{\max}(t), \quad T_{\min}\equiv T_{X_{\min}}.
\label{eq:mainres_app}
\end{equation}
Note that in NESS ($\dot X_{\min}=0$) the solution to \refeq{eq:multiricc} is given by
$X_{\min}^{\text{NESS}}=k T_{\min}^\text{NESS} K^{-1}$, where
$T_{\min}^\text{NESS}$ is given by the same formula as for the overdamped Langevin case, \refeq{eq:TminNESS}, using
$1/R = r$ in the definition of  $\sigma$. In NESS, it holds that the maximum work extraction rate is exactly given by
\begin{equation}
\dot W_{\max}^\text{NESS} = k T_{\min}^\text{NESS} \dot{I}_c^{\text{NESS}},
\end{equation}
and the lower bound in \refeq{eq:mainres_app} is reached.
Therefore, it is only in an initial transient phase where we expect some slack in
the inequality. As $t\rightarrow \infty$, the lower bound approaches an equality, as claimed in the introduction in \refeq{eq:mainres2}.

Finally, let us prove the upper bound $W_{\max}(t)\leq kTI_c(t)$, claimed after \refeq{eq:mainres2}
for the multidimensional case.
For simplicity, and without loss of generality, let us choose coordinates in
the phase space such that $K=I_{2n}$ (the $2n\times 2n$ identity matrix). Then $X_{\min}(0)=kTI_{2n}$, and from
\refeq{eq:multiricc} it follows that $X_{\min}(t)-kTI_{2n}$ is symmetric negative semi-definite for all $t\geq 0$.
Rewriting the maximum work formula in \refeq{eq:multimaxwork}, using that $\text{Tr}(AB)=\text{Tr}(BA)$ for matrices of compatible dimensions, we
have
\begin{equation}
\begin{aligned}
W_{\max}(t) & = \frac{1}{2V_\text{meas}}  \int_0^t B^T X_{\min}^2 B\,dt' \\
& \leq kT \frac{1}{2V_\text{meas}}  \int_0^t B^T X_{\min}B\,dt' \\ & = kT I_c(t).
\end{aligned}
\end{equation}
The inequality follows since $X_{\min}(t)-kTI_{2n}$ is negative semi-definite. This concludes the proof.

\section{Information rate into the transmission line}
\label{sec:inforate}

We study a semi-infinite lossless transmission line of (real) characteristic impedance $Z$ and temperature $T_\text{dem}$. This line is interconnected to an external circuit and as a result carries a random voltage signal $v(t)$ to infinity. We assume that the signal is Gaussian and that its restriction to disjoint frequency bands are independent, which is always the case for white noise, possibly filtered by linear circuits, as is the case in this paper. This allows us to compute energy or information-related quantities over every infinitesimal frequency band $[f,f+df]$ as a separate channel, and then integrate over all frequencies.

It is well known that a signal carrying a bit across a linear transmission line at temperature $T_\text{dem}$ must be of energy $kT_\text{dem} \ln 2$ at least~\cite{shannon48,landauer1996minimal}. This bound is reached in the limit of low rates for a given frequency band, for Gaussian signals, which in our case is the limit of poor measurements, $\sigma \approx 0$. Therefere we deduce that the information rate carried into the line by a (low) power $\dot W$ is
\begin{equation}
\textrm{InfoRate}=\frac{\dot W}{kT_\text{dem}}.
\label{eq:Landauer3}
\end{equation}

More generally, sending information at a rate $dR$ (in nat/s) over a frequency band $df$ into our linear transmission line of temperature $T_\text{dem}$ can only be achieved with a signal whose energy per bit, or power over information rate, is at least~\cite{shannon48}
\begin{equation}
\dot W_\text{1 bit transmission} \geq \frac{e^{dR/df}-1}{dR/df}kT_\text{dem} \ln 2.
\label{eq:shannon}
\end{equation}
Equation~\eqref{eq:shannon} is satisfied with \emph{equality} for a Gaussian signal, as is the case in this paper. This is the energy stored into the memory tape whenever one bit is written. Equivalently, the information rate into the line is
\begin{equation}
\textrm{InfoRate}\equiv \int dR =\int_0^{\infty } \ln\left(1+\frac{\dot W(f)}{kT_\text{dem}}\right) df \leq \frac{\dot W}{kT_\text{dem}},
\label{eq:shannon2}
\end{equation}
where $\dot W(f) df$ is the power contained in the signal restricted to frequency band $[f,f+df]$.

Let us give a direct physical argument for this formula. A signal over a frequency bandwidth $df$ is completely characterized by a sampling frequency $2df$, as stated by the  Nyquist-Shannon sampling theorem. This means that the signal over a time interval $\Delta t$ can be reconstructed in a unique way from just $2df \Delta t$ samples of it. In other words, the signal has exactly $2df \Delta t$ degrees of freedom.

Let us give concrete examples of such degrees of freedom. Those samples can be measured on the signal observed at regularly spaced time instants at a given point of the line. Since the information is travelling along the line as a wave, this can also be achieved by measuring the signal at a given time at regularly spaced points of the line. A common physical model for the transmission line is an infinite ladder of small inductances and capacitances~\cite{cheng}, just as an elastic string is seen as a sequence of small masses and springs. We can therefore measure $2df \Delta t$ currents or voltages in different elements of the line to reconstitute the whole signal.

We choose the degrees of freedom $x_1, x_2, \ldots, x_{2df \Delta t}$ to be uncorrelated and normalized so that the associated energy is $x_i^2/2$. If not for the external signal, each variable $x_i$ would have an energy $kT_\text{dem}/2$, thus a variance $\langle x_i^2 \rangle=kT_\text{dem}$, from the equipartition theorem. Due to the signal, every variable has an extra energy $\frac{\dot W(f) df \Delta t}{2df\Delta t}$, or $\dot W(f)/2$, thus a total variance $kT_\text{dem} + \dot W(f)$. As the differential entropy of a Gaussian variable of variance $V$ is $\frac{1}{2}\ln V + \ln \sqrt{2\pi e}$, this extra variance due to the signal leads to an extra entropy $\frac{1}{2}\ln (kT_\text{dem} + \dot W(f))- \frac{1}{2}\ln kT_\text{dem}$, or $\frac{1}{2}\ln (1+ \frac{\dot W(f)}{kT_\text{dem}})$, on every degree of freedom. Summed over $2df \Delta t$ degrees of freedom, and integrated over all frequencies, we recover Eq. (\ref{eq:shannon2}) above.

This interpretation has a direct connection with Landauer's principle, as follows. Recall that the differential entropy $h(x)$ of a random variable $x$ taking real values with probability density $\phi$ is $-\langle \ln \phi \rangle$. The number of discrete Shannon bits required to encode the value taken by a continuous random variable $x$ with accuracy $\epsilon$ is $(h(x)-\ln \epsilon)/\ln(2)$, in the limit of small $\epsilon$. Therefore the entropy difference $(\frac{1}{2}\ln (kT_\text{dem} + \dot W(f))- \frac{1}{2} \ln kT_\text{dem})/\ln(2)$ is precisely the number of discrete Shannon bits that we would have to remove from every degree of freedom in order to erase the effect of the signal, and restore the line to its original state.

Nevertheless, it is shown in the text that Landauer's bound on erasure cost is tight only in the poor measurements limit, when erasure can be arbitrarily slow as the information rate into the line approaches zero. As the thermodynamic argument in the main text does not make any assumption on the shape of signal, it is valid for any distribution of power $\dot W(f)$ over frequencies. In particular, the implementation cost of the demon to absorb a signal restricted to the frequency band $[f,f+df]$ is
\begin{equation}
\dot W_\text{impl}(f) df=\frac{T}{T_\text{dem}} \dot W(f)df=(e^{dR/df}-1)kT df,
\end{equation}
for an associated information rate $dR$. Therefore we must postulate, in the present context, that the non-equilibrium Landauer per-bit cost for erasing information at rate $dR$ from the frequency band $[f,f+df]$ is at least
\begin{equation}
\dot W_\text{1 bit erasure}=\frac{e^{dR/df}-1}{dR/df} kT  \ln 2
\label{eq:noneqLand}
\end{equation}
to be disposed to a heat bath of temperature $T$. We recover the usual Landauer cost $kT \ln 2$ for the limit of slow rates $dR/df \rightarrow 0$. Erasing a bit on a narrow band at high speed is therefore exponentially more costly then infinitely slow erasure. Note that this non-equilibrium form of Landauer's principle is strongly dependent on our assumptions, mainly linearity of the line. Integrating the cost (\ref{eq:noneqLand}) over all bits at all frequencies, we recover total implementation cost $\dot W_\text{impl}=T \dot W/T_\text{dem}$ as required.

In conclusion, the demon's implementation cost can therefore be entirely attributed to erasure of information in the demon's memory, through Landauer's principle, corrected for the fast rates observed out of equilibrium ($T_\text{min} < T$).

%\bibliography{refs_thermo}
\input{MaxDemon-arXiv-v2.bbl}

\end{document}